\documentstyle[12pt]{article}
\newcommand{\doublespace}{\addtolength{\baselineskip}{.75\baselineskip}}

\begin{document}

\doublespace

\bf{\large{\centerline{\textbf{Formation of antihydrogen in
antiproton - positron collision }}}}
\par
\normalsize\vspace{0.5cm}\centerline{S.Roy, S. Ghosh Deb and C.
Sinha} \centerline{Theoretical Physics Department,}
\centerline{Indian Association for the Cultivation of Science,
Kolkata-700032, India.}
\section{\leftline{Abstract:}}

\par\hspace{1.5 cm}
{\large A quantum mechanical approach is proposed for the
formation of antihydrogen $( \bar{H}) $ in the ground and excited
states (2s, 2p) via the mechanism of three body recombination (
TBR ) inside a trapped plasma of anti proton ( $ \bar{p} $ ) and
positrons ( $ e^{+} $ ) or in the collision between the two beams
of them. Variations of the differential ( DCS ) as well as the
total ( TCS ) formation cross sections are studied as a function
of the incident energies of both the active and the spectator $
e^{+} $s. Significantly large cross sections are found at very low
incident energies in the TBR process as compared to other
processes leading to antihydrogen. The present ( $ \bar{H} $ )
formation cross section decreases with increasing positron energy
( temperature ) but no simple power law could be predicted for it
covering the entire energy range, corroborating the experimental
findings qualitatively. The formation cross sections are found to
be much higher for unequal energies of the two $ e^{+}s $ than for
equal energies, as expected physically.

\section{\leftline{Introduction:}}
\par\hspace{1.5cm}
Production of antihydrogen, the simplest and the most stable bound
state of antimatter is one of the current topics nowadays both
from the experimental and the theoretical perspectives mainly
because its study provides various fundamental differences between
matter and antimatter. Particularly, cold antihydrogen $ (
\bar{H}$ ) atom is an ideal system for studying the fundamental
symmetries in physics e.g., the CPT invariance theorem in the
standard quantum field theory and the gravitational weak
equivalence principle for antimatter. The major challenge facing
the $ \bar{H}$ research is the production of cold and trapped
ground state $ \bar{H}$ that is needed for the precise laser
spectroscopy. Apart from these, there are many important practical
applications of the $ \bar{H} $  out of which the followings are
worthy to be mentioned.
 First, the antihydrogen may also be used for igniting inertial
 confinement fusion pellets, the feasibility of which was
 already  investigated     [1]. Second, the antihydrogen finds important
 applications in the propulsion system [ 2 ].
 \vspace {0.5cm}
\par\hspace{1.6cm} In view of the recent technological advances in the cooling and
trapping mechanism of antiprotons ( $ \bar{p} $ ) and  positrons (
$ e ^{+}$ ), the long term goal for the production of cold and
trapped $ \bar{H} $,
 necessary for the high precision spectroscopic studies has now
 become possible. This has motivated theoretical workers to
 venture different processes producing antihydrogen. The most
 important of these processes is the following three body
 recombination ( TBR ) \hspace{0.2cm}  $ \bar{p} + e^+ + e ^+ \rightarrow\bar{H}+e^+ $
 \hspace{0.3cm} ( I ) in which a spectator particle carries away the excess energy and the  momentum
  released in the recombination. The above reaction poses to be more
  efficient by orders of magnitude [ 3 , 4 ] compared to other $ \bar{H} $ production processes, e.g., the
radiative recombination ( RR ) [ 5 - 7 ], the three body charge
transfer between the Ps and the $ \bar{p}$
   [ 8 - 19 ]. The main reason for this is due to  the following.
   The spectator positron
   in the TBR process efficiently carries off the extra energy, unlike the
   other $ \bar{H} $ production reactions. Another important advantage of the process ( I ) is that the
  reactants are stable charged particles which can be held in a
  trap for cooling and then subsequently for the recombination to
  occur. In fact, it is found experimentally [ 3 , 4  ] that the  TBR in the
trapped plasma of antiproton and positrons happens to be the most
efficient $ \bar{H}\ $ production reaction at low and intermediate
energies. However, the main disadvantage of the TBR is that the $
\bar{H}\ $ is favourably formed in the excited states [ 3 ],
although for the high precision spectroscopic studies, the ground
state $ \bar{H} $ is highly needed. In contrast, in the RR
process, although the ground state is favoured, the cross-section
itself is much lower \newline [ 3 , 4 ] . \vspace{0.5cm}
\par\hspace{1.4cm}
Regarding the experimental situation for the TBR process, the
three main International Groups are working on it with much
endeavour at Cern, e.g., ATHENA [ 20 - 25 ], ATRAP  \newline  [ 26
- 30 ] and Harvard [ 31 - 33 ] while another group from Riken [ 34
, 35 ] is also concentrating on the experiments of cold and
trapped $ \bar{H}\ $ production. In all the experiments attention
is being paid mainly to the temperature dependence of the $
\bar{H}$ production at extreme low energies in the range of meV.

 \vspace{0.5cm}
\par\hspace{1.5cm}
As for the theoretical situation, the first detailed study for the
TBR is due to Robicheaux [  36 ] in the framework of classical
trajectory Monte Carlo ( CTMC ) method. However, in this
calculation the Author introduced some fraction of electrons along
with the $ e^{+}$ as well as a strong magnetic field in order to
make the process feasible. It was noted [ 36 ] that the $\bar{H}$
formation reduces substantially in presence of the $ e^{-}$s.
Prior to and also following this work [36], there exist some
calculations by the same Author [ 37 , 38 ] that mainly study the
temperature dependence of the TBR process based on some
statistical models. \vspace{0.5cm}
\par\hspace {1.5cm}
The present work addresses the study of the  $ \bar{H}\ $
formation cross sections ( both differential and total ) through
the TBR mechanism in the collision between the positron and the $
\bar{p}\ $ plasma. To our knowledge, this is the first quantum
mechanical attempt along this direction in the TBR process.
Although experimentally the TBR process favours the $ \bar{H}\ $
formation in highly excited states, for the theoreticians it is
much easier to calculate the cross sections in the ground and  low
lying states. Thus in the absence of any experimental cross
section data, the present theoretical estimates of the ground and
excited ( 2s , 2p ) states $ \bar{H}$ formation cross-sections
might give some stimulus and guidelines to the future detailed
experiments.
 \vspace{0.5cm}
\par\hspace{1.5cm} The present model corresponds to the following situations .
We consider an ensemble of weakly correlated positrons and the
$e^{+}$ plasma  density is assumed to be low enough so that the $
e^{+} - e^{+}$ interaction can be treated as a perturbation. The
antiproton is treated as a stationary ionic target located at the
origin of coordinates which corresponds to the experimental
situation of a cold and trapped $ \bar{p}\ $. The latter
assumption should be quite legitimate when the positron velocity
is much faster than that of the antiproton which happens to be the
case due to large mass difference  between the two. Since the
recombination reaction requires a third body for the energy and
momentum conservation of the process, another $ e^{+} $ ( electron
) of the plasma serves this purpose and the process becomes a TBR
one. Further, there is a probability of exchange between the
active and  the spectator positrons which is also incorporated in
the present model.
\par\hspace{1.5cm}

\section{\leftline{Theory:}}
\par \hspace{1.5 cm}
 The present study deals with the following three body
 recombination process :
 \begin {center}
$\bar{p}\hspace{.3cm}  + e^+ + e^+ \rightarrow \bar{H}+ e^+ .
$\hspace{5 cm}{( 1 )}\end {center} In the present formulation the
$ \bar{p}$ is assumed to be stationary\\( target ). The prior form
of the transition amplitude $ T_{if} $ for this process is given
by
\begin {center}
$T_{if} = \langle\Psi_{f}^{-}{(
\vec{r}_{1},\vec{r}_{2})}{(1+\textbf{P})}\hspace{.1cm}|\hspace{.1cm}V_{i}\hspace{.1cm}|\hspace{.1cm}\psi_{i}\hspace{.1cm}{(
\vec{r}_{1},\vec{r}_{2})}\rangle$ ,\hspace{1.7 cm}{( 2 )}\end
{center} where \textbf{P} denotes the exchange operator
corresponding to the interchange of the positrons  in the final
channel.
 $V_{i}$ in equation ( 2 ) is the initial channel perturbation which is the
part of the total interaction not diagonalised in the initial
state and $ \psi_{i}$ is the corresponding asymptotic wave
function . The final channel wave function $ \Psi_{f}^{-} $
satisfies the three body Schrodinger equation obeying the incoming
wave boundary condition;  \begin {center} $ ( H - E )\Psi_{f}^{-}=
0 . $ \hspace{6.2 cm}{( 3 )}\end {center} The total Hamiltonian (
H )
 of the system can be written as
\begin {center} $ H =
-\frac{1}{2}\nabla_{1}^{2}\hspace{.1cm}-\frac{1}{2}
\nabla_{2}^{2}\hspace{.1cm}-\frac{1}{r_{1}}\hspace{.1cm}
-\frac{1}{r_{2}}\hspace{.1cm}+\frac{1}{r_{12}},$\hspace{2.8 cm}{(
4 )}\end {center} where ${\vec{r}_1}$ and $ {\vec{r}_2} $
represent the position vectors of the active
 $e^{+}$ ( to be transferred ) and the spectator $e^{+}$ s
 respectively. The atomic unit ( a.u. ) is used
throughout the work.
\par \hspace{1.5 cm}The initial channel asymptotic wave function  $
\psi_{i}$ in equation ( 2 ) satisfies the following Schrodinger
equation  \begin {center} $(
-\frac{1}{2}\nabla_{1}^{2}\hspace{.1cm}-\frac{1}{2}\nabla_{2}^{2}
\hspace{.1cm}-\frac{1}{r_{1}}\hspace{.1cm}- \frac{1}{r_{2}}-E
\hspace{.2cm}) \hspace{.2cm}\psi_{i}= 0 $\hspace{4.6 cm}{(5)}\end
{center} and is given by
\begin {center} $ \psi_{i}= N_{j} \ e^{i\vec{k}_{j}\cdot
\vec{r}_{j}}\hspace{0.2 cm}_1F_{1}\hspace{.2
cm}[\hspace{.2cm}i\alpha_j,1,
 -i\hspace{.2cm}(k_j r_j -\vec k_j\cdot\vec r_j)],\hspace{.2
 cm}$\hspace{2.5
cm}{(6)}\end {center} with $ N_{j}= \exp ( \frac{-\pi
\alpha_{j}}{2} )\hspace{.1cm} \Gamma ( 1 - i \alpha_{j}) ; $ j =
1, 2 ; $ \alpha_{j}= - \frac{1}{k_{j}}$ ; $ \vec{k}_{j}$ denotes
the incident momentum of the active or the spectator $ e^{+}$
respectively. The approximated final state wave function
$\Psi_{f}^{-}$ is chosen in the framework of eikonal approximation
as follows :\begin {center} $ \Psi_{f}^{-}=
\phi_{f}(r_{1})\hspace{.1 cm}
e^{i\vec{k}_{f}\hspace{.03cm}\cdot\hspace{.03cm}\vec{r}_{2}}
\hspace{.1cm}\exp \hspace{.1 cm}[ i
\eta_{f}\hspace{.1cm}\int_{z}^{\infty}\hspace{.1cm}(
\frac{1}{r}_{12}- \frac{1}{r}_{2})dz'] $\hspace{2.7 cm}{(7)}\end
{center}with $ \eta_{f}=\frac{1}{k_{f}}$, $ k_{f} $ being the
final momentum of the spectator $ e^{+} $; $\phi_{f}({r_{1}})$
represents the bound state wave function of the $\bar{H}$ atom .
\par \hspace{1.5cm}
Finally the differential cross section for the process \newline( 1
) is given by \begin {center}$ \frac
{d\sigma}{d\Omega}=\frac{k_{f}}{k_{1}
k_{2}}\hspace{.2cm}[\frac{1}{4} \hspace{.03cm}( \hspace{.05cm}| f
+ g |^{2} \hspace{.03cm}) + \frac{3}{4} (\hspace{.05cm} | f - g
|^{2} \hspace{.03cm})] , $ \hspace{4.2 cm}(8)\end {center}\
\newline where f and g corresponds to the direct and the exchange
amplitudes respectively.

\par \hspace{1.5 cm} Using the following contour integral representations of the
eikonal phase factors [ 39 ] as well as the coulomb functions
\newline [ 40 ] and after much analytical reductions  [ 41, 42  ], the
transition matrix element ( 2 ) is finally reduced [ 43 ] to a
three dimensional integral which is evaluated numerically by using
different quadrature methods . The eikonal phase factor is of the
form
\begin {center} $ y\pm^{( i \eta - n )} = \frac{(-1)^{n+1}}{2i \sin(\mp \pi i
\eta)\hspace{.2cm}\Gamma(\mp i \eta\pm n )} \int _{c}( - \lambda )
^{\mp i \eta\pm n - 1 }\exp(-\lambda y ) dy $ \hspace{.4 cm}{( 9a
)}\end {center} where the contour c has a branch cut from 0 to
$\infty$ [ 39 ] ; the confluent hypergeometric function:\ \begin
{center} $ _1F_{1}\hspace{.2 cm}( i\alpha,1, z ) = \frac{1}{2\pi
i}\hspace{.1cm}\int_{\Gamma_{1}}^{(0^{+}, 1^{+})}\hspace{.1cm}
 p \hspace {.1cm}( \alpha, t ) \exp ( z t ) dt \hspace{.2 cm}$\hspace{3
cm}{( 9b )}\end {center}  with $ p \hspace {.1cm}( \alpha, t ) =
\hspace {.1cm} t ^{-1 + i \alpha } ( t - 1 )^{- i \alpha}$, $
\Gamma _{1}$ is a closed contour encircling the two points 0 and 1
once anti - clockwise [40]. At the point where the contour crosses
the real axis to the right side of 1,  arg $\emph{t}$ and arg
($\emph{ t-1  }$)  are both zero.
\section{\leftline{ Results and Discussions:}}
\par \hspace{1.5 cm} We have computed the $ \bar{H} $ formation
cross sections both differential and total for the TBR process
  ( 1 ) in the framework of the coulomb distorted eikonal
approximation ( CDEA ), where distortions have been included in
both the channels. The exchange between the active and the
spectator $ e^{+} $s is also incorporated. Since the present
process ( 1 ) is an exothermic reaction it can occur even at zero
incident energy. However, our results are not converged below 5 eV
due to computational problems and are therefore not reported here.
Furthermore, it may be mentioned that the present model might not
yield very reliable results at extreme low energies.  \vspace{.5
cm }
\par\hspace{1.5cm}Figs ( 1 - 5 ) exhibit the present differential
cross sections ( DCS ) in the ground and excited states ( 2s and
2p ) for different incident energies of both the positrons e.g., $
E_{1} = E_{2}= $ 10, 20, 25, 30 and 50 eV respectively. The
figures reveal that the $ \bar{H} $ formation ( in all the states
) is strongly favoured in the forward directions and as such the
DCS are presented upto $
 60^{0} $  only, beyond which the cross sections become
negligible. At very low incident energy, the magnitude of the
formation cross section is found to be largest for the 2p state
and smallest for the 1s state while the 2s lies in between, i.e.,
 $ 2p> 2s> 1s $ ( vide fig. 1 ). This trend of the DCS is noted
upto 15 eV  ( not shown in figure), although with increasing
incident energy, the maximum of the 2p DCS decreases and tends
towards the 2s maximum so that at $ E_{1} $ = $ E_{2}$ = 20 eV,
the 2s overtakes the 2p ( vide fig. 2 ). The DCS peak in this case
is in the order $ 2s> 2p> 1s $. In contrast, at intermediate and
high incident energies ( $ \sim 25  $  eV onwards ), the DCS is
maximum for the 1s state and minimum for the 2p state while the 2s
lies in between ( i.e. $ 1s> 2s> 2p $, figs. 3 - 5 ).\vspace{.5cm}
\par\hspace{1.5cm} As for the position of the DCS maximum, at low
incident energies ( $\sim$ upto 15 eV ), the 1s and 2p maxima lie
at some lower scattering angles (  $\sim 20\hspace{.05cm}^{0}  $ )
while the 2s maximum occurs at the extreme forward ( $ \sim 0
\hspace{.05cm}^ {0}$ ). With increasing energy, the DCS maxima for
these two states ( 1s and 2p ) move towards the extreme forward (
$ \sim  0 \hspace{.05cm}^ {0}$ ), while the 2s maximum moves in
the reverse direction ( vide figs. 2 - 4 ). However, at high
incident energies (e. g., $\sim $ 40 eV onwards ), all the partial
DCS maxima are finally peaked at extreme forward $ 0 ^{0}$ ( vide
fig. 5 ), as expected. \vspace{.5cm}
\par \hspace{1.5cm}
Figs. 6(a) - 6(c) again exhibit the partial DCS but for some
unequal energies of the two incident $ e^{+} $ s ( i.e. $
E_{1}\neq E_{2} $ ) along with a case for $E_{1}=E_{2}$ ( 15 eV )
for the sake of comparison. The following interesting features are
noted from the figures. All the partial DCS are found to be much
higher ( by a factor of $\sim$ 2 to 2.5 ) when the energy of the
active $e^{+}$  ( $ E_{1} $ ) is greater than that of the
spectator one ( $ E_{2}$ ), i. e., when $ E _{1}
> E_{2}$. The DCS  for unequal ( $ E_{1} \neq E_{2} $ )
energies lie much above than those for equal energies  ( $ E_{1} =
E_{2} $ ). This is quite expected physically due to strong
repulsion between the two $ e^{+} $ s at equal energies.
\vspace{.5cm}
\par \hspace{1.5cm}
Next we come over to the total cross sections  ( TCS ) for the $
\bar{H} $ formation displayed in
 figs. 7  and 8 for different sets of incident
 energies. Figure 7 displays the partial TCS  when the two incident $ e^{+} $s share
equal energy $( E_{1}= E_{2} )$ . As in the case of DCS, at low
and intermediate incident energies ($ \sim $ upto 20 eV), the
partial TCS follows the order $ 2p > 2s > 1s $ ( inset of fig.7 ),
while beyond
 25 eV it is in the decreasing order with excitation of
the  $ \bar{H} $ state, i.e.,
  $ 1s > 2s > 2p $. The dominance of the 2p TCS at low incident energies
  could probably be attributed to the long range polarization effects which is
  much stronger for the 2p state than for any other states. In fact, a major
  contribution to the polarization effect that mainly dominates at
  lower incident energies, comes from the lowest lying p state ( i.e., 2p state ).
   \vspace{.5cm}
\par \hspace{1.5cm}Fig. 7 also indicates that although all the partial TCS
 decrease monotonically with increasing incident energy, they
 do not follow any simple power law ( e.g. $ \sim  T^{-9/2}$ ),
 corroborating the experimental findings [ 21 , 23 ] .
\vspace{.5cm}
\par \hspace{1.5cm} Figures 8(a) - 8(c) display
the partial TCS against the active $ e^{+} $ energy ( $ E_{1} $ )
for some fixed values of $ E_{2} $ ( spectator ) while the insets
exhibit the reverse one, i.e. TCS vs $ E_{2} $ for fixed $ E_{1}
$. As in the case of DCS, for a fixed sum of $ E_{1} $ and $ E_{2}
$, the partial TCS is found to be larger when $ E_{1} > E_{2} $
than for $ E_{1} < E_{2} $. Further, the
 TCS against $ E _{2} $  falls off
much more rapidly than the TCS versus $ E _{1} $ ( cf. figs. 8
with their insets ). Regarding the relative magnitude of the
partial TCS, for lower energy of the spectator $  e^{+} $, e. g.,
$ E_{2} $ = 10 eV ( fig. 8(a)), the general trend of the TCS
follows the order $ 2p > 2s > 1s $ at low and intermediate  $
E_{1}$ while at higher $ E_{1} $, the above order changes to $ 2s
> 2p > 1s $. Similar behaviour is noted in fig. 8(b)( for $
E_2$ = 15 eV ) as in fig. 8 (a) with some exceptions at higher $
E_1$ .  At intermediate $ E_2$ (25 eV, fig. 8 (c)), the partial
TCS follow different orders for different ranges of $ E_1$, e.g.,
at lower $ E_1$ the 2p dominates while at higher $ E_1$ the 1s
dominates. However, at higher  $ E_2$ ( $\sim E_2 \geq $ 50 eV ),
the 1s cross section dominates through out the range of $ E_1$
except at very low energies ( $ E_1 \sim 5 - 10$eV ) where the 2s
is most prominent ( not shown in figure ). \vspace{.5cm}
\par \hspace{1.5cm} Figure 9 demonstrates a comparison  between
the present TCS and the corresponding experimental results of
Merrison et al [ 44 ] for the $ Ps ( 1s ) + \bar{p}
\longrightarrow \bar{H} ( 1s ) + e ^{+} $  process. The
experimental Ps energies are converted to the antiproton energies
following the relation : $ E _{\bar{p}} ( Kev ) =
\frac{K_{Ps}^{2}* 6.8 * 918 }{1000} $. As was anticipated [ 3 , 4
], the present TBR cross sections are found to lie much above the
experimental data [44] for the abovesaid process.\vspace{.5cm}
\par \hspace{1.5cm} For the sake of some numerical measures, we have
displayed in Table I the present partial ( 1s, 2s, 2p ) TCS along
with some other existing theoretical results due to Mitroy et al
\newline [ 14, 17 ] for the process $ \bar{p} + Ps \longrightarrow \bar{H}(
n, l, m ) + e^{+} $ using unitarized Born approximation [ 17 ] and
close coupling approximation [ 14 ]. Results due to Sinha et al [
19 ] for the above process in the eikonal approximation both with
and without laser field are also included in the Table I. The
incident energies are chosen in accordance with their [ 14 , 17 ]
calculations. The field assisted \newline ( FA ) results [ 19 ]
are presented for the field strength 0.01 a.u. and the frequency
0.043 a.u.}
\begin{center} Table I
\end{center}
\small
\begin{tabular}{|c|c|c|c|c|}
  \hline
  Energy & Mitroy et al [14 ]  & Mitroy et al [17 ]  & Sinha et al [ 19 ] & Present results\\

    ( eV )  &  1s\hspace{.5cm}( 2s+2p) &   1s \hspace{.5cm}2s\hspace{1cm}2p&FF ( 1s )\hspace{.5cm}FA( 1s )
 &1s \hspace{.2cm}2s\hspace{.2cm} 2p\\
   \hline
   13.60 & \hspace{.1cm}1.923 \hspace{.5cm} 8.44&\hspace{0.2cm}1.46 \hspace{.4cm}0.66 \hspace{.2cm} 3.81 & 0.92\hspace{.4cm} 8.32 & \hspace{0.5cm}373.07 \hspace{0.2cm} 705.4\hspace{0.3cm}  530.06  \\
   \hline
   20.40 & -  \hspace{0.2cm} &\hspace{0.2cm}0.94 \hspace{.2cm} 0.254 \hspace{.2cm} 1.76 & 0.47\hspace{.4cm}3.99 &\hspace{0.5cm} 68.25 \hspace{.4cm} 59.62 \hspace{.2cm} 36.13 \\
   \hline
   25.84 &\hspace{0.1cm} 0.735 \hspace{.5cm} 1.729 & - & 0.29\hspace{.4cm}2.05& \hspace{0.8cm}23.16 \hspace{.4cm} 15.43 \hspace{.2cm} 11.048 \\
   \hline
   34.00 & -   & \hspace{0.1cm}0.394 \hspace{.1cm}0.08 \hspace{.2cm} 0.396 & 0.16\hspace{.4cm}0.65 & \hspace{0.7cm}5.14 \hspace{.6cm} 3.29 \hspace{.6cm} 1.21 \\
   \hline
   43.52 &  0.24 \hspace{.7cm} 0.28 & -  & 0.08\hspace{.4cm}0.19 & \hspace{0.5cm}1.38 \hspace{.6cm} 0.73 \hspace{.6cm}  0.21 \\
   \hline
  54.40& -   & 0.13 \hspace{.2cm} 0.03 \hspace{.2cm} 0.06 & 0.04\hspace{.4cm}0.09&\hspace{0.5cm} 0.507 \hspace{.3cm} 0.212 \hspace{.3cm} 0.651 \\
  \hline
   63.92 &  0.078  \hspace{.5cm}  0.053   &     -      & 0.02\hspace{.4cm}0.05& \hspace{0.6cm}0.165 \hspace{.4cm} 0.064 \hspace{.2cm} 0.034 \\
\hline
\end{tabular}
\vspace{.5cm}
{\large
\par \hspace{1.5cm} Table I again confirms ( as in fig.9 ) that the present TBR
cross sections are much larger than all the other processes
leading to antihydrogen throughout the energy range
considered.\vspace{.5cm}\newpage
   \par\par \hspace{1.5cm}
  Table II displays the probable power laws obeyed by  the
 partial as well as the sum TCS
 for different incident energy ranges of the $ e^{+} $
 corresponding to figures 7 ( $ E_{1}=E_{2} $  ) and 8 ($ E_{1}\neq E_{2}$).\\

 \begin{center} Table II
 \end{center}

\begin{tabular}{|c|c|c|c|c|}\hline
\small
Energy Range & \multicolumn{4}{c|}{Power law obeyed}\\
( in eV )  &   \multicolumn{4}{c|}{}  \\
  \hline

\ $ E_{1}=E_{2} $    & \bf1s &\bf 2s &\bf 2p &\bf 1s + 2s + 2p \\
\par\hline
\vspace{0.1cm}
  5 - 10  & $\sim E ^{-2.9}$ & $\sim E^{-3.7}$ & $\sim ~E^{-4.8}$ & $\sim ~E^{-4.5}$
  \\

  10 - 25  &$\sim E^{-3.9}$& $\sim E^{-5.3}$ & $\sim E^{-6.0}$ & $\sim E^{-5.4}$
  \\

  25 - 50 & $ \sim E^{-5}$ &$~\sim E^{-6.1}$ & $\sim E^{-6.7}$ & $\sim E^{-5.9}$
  \\
  \hline
 $E_{1} \leq E_{2}$ ( 10 eV ) & \multicolumn{4}{c|}{}\\
  {}  &   \multicolumn{4}{c|}{}  \\
 \hline

    5 - 10 &  $\sim E_{1}\hspace{0.02cm}^{- 1.6}$ &  $\sim E_{1}\hspace{0.02cm}^{- 1.3}$ &  $\sim E_{1}\hspace{0.02cm}^{- 1.8}$&  $\sim E_{1}\hspace{0.02cm}^{- 1.6}$
\\ {}  &   {} &{} &{}&{}  \\
\hline

  $E_{1} \geq E_{2}$ ( 10 eV )& \multicolumn{4}{c|}{}\\
   {}  &   \multicolumn{4}{c|}{}  \\
\hline
    10 - 25 &  $ \sim E_{1}\hspace{0.02cm}^{- 1.5}$ &  $\sim E_{1}\hspace{0.02cm}^{- 1.7}$ &  $\sim E_{1}\hspace{0.02cm}^{- 2.1}$&  $\sim E_{1}\hspace{0.02cm}^{- 1.9}$\\
    25 - 50 &  $\sim E_{1}\hspace{0.02cm}^{- 1.5}$ &  $\sim E_{1}\hspace{0.02cm}^{- 1.8}$ &  $\sim E_{1}\hspace{0.02cm}^{- 2.3}$&  $\sim E_{1}\hspace{0.02cm}^{- 1.9}$\\
    \hline
\end{tabular}
 \vspace{0.5cm}

\vspace{0.5cm}
\par \hspace {1.5cm} As is revealed from the table , the low energy partial TCS
( e.g. , $E_{1}=E_{2}\sim$ 5 - 10 eV ) falls off much slowly as
compared to the intermediate and high  energies and the slope of
the 1s TCS ( see also fig. 7 ) is much less than that of the
others
\newline ( 2s, 2p and 1s + 2s + 2p ). Another important feature should be
noted from this Table that for $ E_{1}\neq E_{2}$, the power  of
the exponent decreases as compared to the $E_{1}= E_{2}$ case
throughout the energy range. This again indicates the better
efficiency of the $ \bar{H}$ production for unequal energies ( $
E_{1}\neq E_{2}$ ) of the active and the passive $ e^{+}$ s over a
wider energy ranges ( $ E_{1} $).

\section {\leftline {Conclusions:}}
\par \hspace{1.5 cm} The salient features of the present study are
as follows:\vspace{0.5cm}
 \par \hspace{1.5 cm}At very low incident energies the present TBR cross
section for the $ \bar{H}$ formation in the 2p state is found to
be the dominant process among the three states 1s, 2s, 2p while at
intermediate and high incident energies, the ground state ( 1s )
cross section dominates for both equal and unequal energies of the
two positrons with some exceptions for the latter case (
$E_{1}>E_{2}$). \vspace{0.5cm}
\par \hspace{1.5 cm} Substantially high cross sections
 are noted in the TBR model than in the other RR / charge
 transfer processes leading to antihydrogen.
\vspace{0.5cm}\par \hspace{1.5cm} The partial TCS is found to be
significantly higher when the active $ e^{+}$ energy is greater
than that of the spectator $ e^{+}$ ( i. e., $ E_{1}> E_{2} $)
than for  $ E_{1}=E_{2}$ or for  $ E_{1}< E_{2}$.\vspace{0.5cm}
\par\hspace{1.5cm}
For a more efficient production of $ \bar{H} $  for a wider energy
range, the unequal ( $ E_{1} > E _{2 } $ ) distribution of energy
between the active  and the spectator positrons could be suggested
rather than the equal one ( $ E_{1} = E_{2} $ ). \vspace{0.5cm}
\par \hspace{1.5cm}
The present $ \bar{H} $ formation cross section decreases with
increasing $ e^{+} $ energy ( i.e., temperature ) but does not
follow any simple scaling law ( e.g., $ \sim  T ^{-9/2} $ ),
corroborating the experimental findings. However, both the partial
and the sum TCS obey different power laws for different incident
energy ranges. \vspace{0.5cm}

\par\hspace{1.5 cm} Finally, the present results might be important
for the future detailed $ \bar{H} $ experiments.
\section{\leftline{ References:}}
\lbrack 1\rbrack \hspace{0.2cm} Andre Gsponer and Jean - Pierr
Hurni, arXiv: physics/0507125v2 [ physics.plasm-ph ] 17 July,
 ( 2005 ).\\\\
\lbrack 2\rbrack \hspace{0.2cm} Special issue of the Journal of
the British Interplanetary Society on antimatter propulsion. JBIS
\bf{35} \normalsize ( 1982 ) 387. - R. L. Forward : \emph{Making
and storing antihydrogen for propulsion, } Workshop on the Design
of a low Energy Antimatter Facility in the USA , University of
Wisconsin, October 3 - 5 ( 1985 ) .\\\\
\lbrack 3\rbrack \hspace{0.2cm} G. Gabrielse, S. L. Rolston, L.
Haarsma and W. Kells, Phys.Letters A, \bf{129}\normalsize, 38 (1988).\\\\
\lbrack 4\rbrack \hspace{0.2cm} G. Gabrielse, Advances in Atomic,
Molecular and Optical Physics, \bf{50}\normalsize, 155
(2005).\\\\
\lbrack 5\rbrack \hspace{0.2cm} S. M. Li, Z.J. Chen, Q.Q. Wang,
Z.F. Zhou, Eur. Phys. J. D.,
\bf{7}\normalsize, 39 (1999).\\ \\
\lbrack 6\rbrack\hspace{0.2cm} Shu - Min Li, Yan - Gang Miao, Zi -
Fang Zhou, Ji Chen, Yao - Yang Liu, Phys. Rev. A, \bf{58}\normalsize, 2615 (1998).\\ \\
\lbrack 7 \rbrack\hspace{0.2cm} Saverio Bivona, Riccardo Burlon,
Gaetano Ferrante and Claudio Leone, Optics Express,
\bf{14}\normalsize, 3715 (2006).\\\\
\lbrack 8\rbrack \hspace{0.2cm} J. W. Humberston, M. Charlton,
F.M. Jacobsen, B.I. Deutch, J. Phys. B, \bf{20}\normalsize,
L25 (1987).\\\\
\lbrack 9\rbrack \hspace{0.2cm} J. W. Darewych, J. Phys. B,
\bf{20}\normalsize, 5917
(1987).\\\\
\lbrack 10\rbrack \hspace{0.2cm} S. N. Nahar and J. M. Wadehra,
Phys. Rev. A, \bf{37}\normalsize, 4118, (1988). \\\\
\lbrack 11\rbrack \hspace{0.2cm} M. Charlton, Phys. Lett. A,
\bf{143}\normalsize, 143 (1990).\\\\
\lbrack 12\rbrack\hspace{0.2cm} S. Tripathi, C.Sinha and N. C.
Sil, Phys. Rev. A, \bf{42}\normalsize, 1785 (1990). \\\\
\lbrack 13\rbrack\hspace{0.2cm} J. Mitroy and A. T. Stelbovics,
Phys. Rev. Lett., \bf{72}\normalsize, 3495 (1994).\\\\
\lbrack 14\rbrack\hspace{0.2cm} J. Mitroy and G. Ryzhikh, J.
Phys. B.,\bf{30}\normalsize, L371 (1997).\\\\
\lbrack 15\rbrack\hspace{0.2cm} S. Tripathi, R. Biswas and C.
Sinha, Phys. Rev. A, \bf{51} \normalsize, 3584 (1995). \\\\
\lbrack 16\rbrack\hspace{0.2cm} J. Mitroy, Phys. Rev. A, \bf{52}
\normalsize, 2859 (1995).\\\\
\lbrack 17\rbrack\hspace{0.2cm} J. Mitroy and A. T. Stelbovics,
J. Phys. B, \bf{27}\normalsize, L79 (1994).\\\\
\lbrack 18\rbrack\hspace{0.2cm} D. B. Cassidy, J. P. Merrison, M.
Charlton, J. Mitroy and G. Ryzhikh, J. Phys. B, \bf{32}\normalsize, 1923 (1999).\\\\
\lbrack 19\rbrack\hspace{0.2cm} A. Chattopadhyay, C. Sinha,
Phys. Rev. A,  \bf{74}\normalsize, 022501 (2006).\\\\
\lbrack 20\rbrack\hspace{0.2cm} M.Amoretti et al, Nature, \bf{419}\normalsize, 456 (2002).\\\\
\lbrack 21\rbrack\hspace{0.2cm} M. Amoretti  et al, Phys. Rev .
Lett., \bf{91}\normalsize, 055001 (2003).\\\\
\lbrack 22\rbrack\hspace{0.2cm} M . Amoretti et al, Phys. lett .
B., \bf{578}\normalsize, 23 (2004).\\\\
\lbrack 23\rbrack\hspace{0.2cm} M . Amoretti et al, Phys. lett .
B., \bf{583}\normalsize, 59 (2004).\\\\
\lbrack 24\rbrack\hspace{0.2cm} N. Madsen et al, Phys. Rev .
Lett., \bf{94}\normalsize, 033403 (2005).\\\\
\lbrack 25\rbrack\hspace{0.2cm}M. Amoretti  et al, Phys. Rev .
Lett., \bf{97}\normalsize, 213401 (2006).\\\\
\lbrack 26\rbrack\hspace{0.2cm} G. Gabrielse et al, Phys. Rev.
Lett., \bf{89}\normalsize, 213401 (2002).\\\\
\lbrack 27\rbrack\hspace{0.2cm} G. Gabrielse et al, Phys. Rev.
Lett., \bf{89}\normalsize, 233401 (2002).\\\\
\lbrack 28\rbrack\hspace{0.2cm} G. Gabrielse et al, Phys. Lett.
B., \bf{548}\normalsize, 140 (2002).\\\\
\lbrack 29\rbrack\hspace{0.2cm} G. Gabrielse et al, Phys. Rev.
Lett., \bf{93}\normalsize, 073401 (2004).\\\\
\lbrack 30\rbrack\hspace{0.2cm} G. Gabrielse et al, Phys. Rev.
Lett., \bf{98}\normalsize, 113002 (2007).\\\\
\lbrack 31\rbrack\hspace{0.2cm} G. Gabrielse et al, Phys. Lett.
B., \bf{507}\normalsize, 1 (2001).\\\\
\lbrack 32\rbrack\hspace{0.2cm} A. Speck et al, Phys. Lett. B,
\bf{597}\normalsize, 257 (2004). \\\\
\lbrack 33\rbrack\hspace{0.2cm} T. Pohl et al, Phys. Rev. Lett.,
\bf{97}\normalsize, 143401 (2006).\\\\
\lbrack 34\rbrack\hspace{0.2cm} G. Andresen et al , Phys. Rev.
Lett., \bf{98}\normalsize, 023402 (2007).\\\\
\lbrack 35\rbrack\hspace{0.2cm} A Torii Hiroyuki et al. Private communication. \\\\
\lbrack 36\rbrack\hspace{0.2cm} F. Robicheaux, Phys. Rev. A
,\bf{70}\normalsize, 022510 (2004).\\\\
\lbrack 37\rbrack\hspace{0.2cm} F. Robicheaux, Phys. Rev. A,
,\bf{73}\normalsize, 033401 (2006).\\\\
\lbrack 38\rbrack\hspace{0.2cm} F. Robicheaux, J. Phys. B,
 \bf{40}\normalsize , 271 (2007).\\\\
 \lbrack 39\rbrack\hspace{0.2cm} I. S. Gradshteyn and I. M. Ryzhik ; Tables and Integrals,
 Series and Products ( Academic New York, 1980 )p. 933. \\\\
 \lbrack 40\rbrack\hspace{0.2cm} A. Messiah ; \emph{Quantum Mechanics} (North - Holland,
  Amsterdam, 1966), Vol. 1, p.481. \\\\
\lbrack 41\rbrack\hspace{0.2cm} R. Biswas and C. Sinha, Phys. Rev.
A ,\bf{50}\normalsize, 354 (1994).\\\\
 \lbrack 42\rbrack\hspace{0.2cm} B. Nath and C. Sinha, Eur. Phys.
 J. D., \bf{6}\normalsize, 295 (1999).\\\\
  \lbrack 43\rbrack\hspace{0.2cm} B. Nath and C. Sinha, J. Phys.
 B., \bf{33}\normalsize, 5525 (2000).\\\\
\lbrack 44\rbrack\hspace{0.2cm} J. P. Merrison, H. Bluhme, J.
Chevallier, B. I. Deutch , P. Hvelplund, L. V. Jorgensen, H.
Knudsen, M. R. Poulsen and M. Charlton, Phys. Rev. Lett.,
\bf{78}\normalsize, 2728 (1997).\\\\ \vspace{0.5cm}

 \end{document}